\documentclass{PoS}
\usepackage{amsmath,amssymb}
\usepackage{bm}
\usepackage{mathrsfs}
\usepackage{cleveref}
\usepackage{braket}

\newcommand{\tr}{\operatorname{tr}}
\crefname{equation}{Eq.}{Eqs.}
\Crefname{equation}{Equation}{Equations}
\crefname{figure}{Fig.}{Figs.}
\Crefname{figure}{Figure}{Figures}
\crefname{section}{Sec.}{Secs.}
\Crefname{section}{Section}{Sections}

\title{
\begin{picture}(0,0)(0,0)%
  \put(360,145){\makebox(0,0)[l]{\textnormal{\normalsize
  CHIBA-EP-233
  }}}%
\end{picture}%
\begin{picture}(0,0)(0,0)%
  \put(360,135){\makebox(0,0)[l]{\textnormal{\normalsize
  KEK Preprint 2018-79
  }}}%
\end{picture}%
Correct way to extract dominant part of the Wilson loop in higher representations 
}

\ShortTitle{Correct way to extract dominant part of the Wilson loop in higher representations}

\author{\speaker{Ryutaro Matsudo}\\
Department of Physics, 
Faculty of Science and Engineering, 
Chiba University, Chiba 263-8522, Japan\\
        E-mail: \email{afca3071@chiba-u.jp}}

\author{Akihiro Shibata\\
Computing Research Center, High Energy Accelerator Research Organization (KEK) \\
SOKENDAI (The Graduate University for Advanced Studies), Tsukuba 305-0801, Japan\\
        E-mail: \email{akihiro.shibata@kek.jp}}

\author{Seikou Kato\\
Oyama National College of Technology, Oyama, Tochigi 323-0806, Japan \\
        E-mail: \email{skato@oyama-ct.ac.jp}}

\author{Kei-Ichi Kondo\\
Department of Physics,  
Graduate School of Science, 
Chiba University, Chiba 263-8522, Japan\\
        E-mail: \email{kondok@faculty.chiba-u.jp}}

\abstract{
The Abelian dominance for the string tension was shown for the fundamental sources in MA gauge in the lattice simulations. For higher representations, however, it is also known that the naive ``Abelian'' Wilson loop, which is defined by using the diagonal part of the gauge field, does not reproduce the correct behavior. To solve this problem, for an arbitrary representation of an arbitrary compact gauge group, we propose to redefine the ``Abelian'' Wilson loop. By using this redefined operator, we demonstrate the ``Abelian'' dominance for sources in the adjoint representation and the sextet representation of $SU(3)$ gauge group in lattice simulations.
}

\FullConference{XIII Quark Confinement and the Hadron Spectrum - Confinement2018\\
		31 July - 6 August 2018\\
		Maynooth University, Ireland}

\begin{document}

\section{Introduction}
The mechanism of quark confinement is not yet satisfactorily understood.
The origin of this non-perturbative feature could be attributed to topological configurations in Yang-Mills theory.
Among such configurations, magnetic monopoles defined by using the Abelian projection is one of the most promising candidates for degrees of freedom responsible for confinement.

In gauge theories with adjoint Higgs field, there are topological soliton solutions which represent magnetic monopoles if the Higgs field acquires a vacuum expectation value and the gauge symmetry $G$ spontaneously breaks to the Cartan subgroup $H$,
where the existence of the topological charge is suggested from the homotopy group $\pi_1(G/H)\neq 0$.
However, in a gauge theory without scalar fields, there are no such solutions of the classical equation of motion, and therefore the definition of monopoles is not clear.
In \cite{Hooft81}, 't Hooft successfully defined monopoles by partially fixing the gauge so that the Cartan subgroup $H$ remains unbroken.
In this method, the ``diagonal component'' of the Yang-Mills gauge field is identified with the Abelian gauge field and a monopole is defined as a Dirac monopole.

It has been tested numerically that these monopoles actually contribute to quark confinement, i.e., the area law of the Wilson loop average.
The numerical simulations have been performed under the Abelian projection using the MA gauge in $SU(2)$ and $SU(3)$ gauge theories on the lattice.
Then it was confirmed that (i) the diagonal part extracted from the original gauge field in the MA gauge reproduces the full string tension calculated from the original Wilson loop average \cite{SY90,STW02}, 
and that (ii) the monopole part extracted from the diagonal part of the gauge field mostly reproduces the full string tension \cite{SS94,STW02}. 

However, it is known that, if the Abelian projection is naively applied to the Wilson loop in higher representations, the resulting monopole contribution does not reproduce the original Wilson loop average \cite{DFGO96}.
This is because, in higher representations, the diagonal part of the Wilson loop does not behave in the same way as the original Wilson loop.
Poulis heuristically found the correct way to extend the Abelian projection approach for the adjoint representation of $SU(2)$ \cite{Poulis96}.

In this talk, we propose a systematic prescription to extract the ``dominant'' part of the Wilson loop average, which can be applied to the Wilson loop operator in an arbitrary representation of an arbitrary compact gauge group.
In order to test this proposal, we calculate numerically the ``dominant'' part of the Wilson loop for the adjoint representation of $SU(2)$ group,  and adjoint and sextet representations of $SU(3)$ group. 
The results support our claim.

\section{Abelian projection} \label{sec:Abel}
In this study, instead of the Abelian projection, we use another way to extract the ``dominant'' part, the field decomposition (for review, see \cite{KKSS15}), which is equivalent to the Abelian projection but explicitly gauge invariant.
For simplicity we use the language of the Abelian projection in this talk, which can be easily translated into the language of the field decomposition.

Now we give a brief review of the Abelian projection.
In this method, we partially fix the gauge so that the residual symmetry group is the Cartan subgroup $H$.
We identify the component $a_\mu$ of the gauge field $A_\mu$ belonging to $H$ with the ``Abelian'' gauge field, and a monopole is defined as a Dirac monopole for this ``Abelian'' gauge field.

In the lattice gauge theory, the monopole contribution to the Wilson loop is extracted as follows.
We put all the configurations into a specific gauge, e.g., the MA gauge, by maximizing a functional for a given configuration $\{U_{x,\mu}\}$ with respect to the gauge transformation, $U_{x,\mu} \rightarrow U^{\Theta}_{x,\mu} := \Theta_x U_{x,\mu} \Theta^\dag_{x+\mu}$.
For the MA gauge, the functional is given as
\begin{align}
  R_{\mathrm{MA}}[U^\Theta] = \sum_{x,\mu}\tr\left(\sum_{a=1}^r H_aU^{\Theta}_{x,\mu}H_a U^{\Theta\dag}_{x,\mu}\right) \label{MA}
\end{align}
where $r$ is the rank of the group and $H_a$ is a Cartan generator $(a=1,\ldots,r)$.
In the present study for the $SU(3)$ gauge theory, 
we performed numerical simulations under two additional gauge conditions which are defined by maximizing the functionals
\begin{align}
  R_{n3}[U^\Theta] &= \sum_{x,\mu} \tr(T^3U^{\Theta}_{x,\mu}T^3U^{\Theta\dag}_{x,\mu}),
\label{n3}
\\
  R_{n8}[U^\Theta] &= \sum_{x,\mu}\tr(T^8U^{\Theta}_{x,\mu}T^8U^{\Theta\dag}_{x,\mu}).
\label{n8}
\end{align}
The Abelian link variable $u_{x,\mu}\in H$ is chosen so as to maximize $\operatorname{Re}\tr(u_{x,\mu}U^{\Theta\dag}_{x,\mu})$.
The Abelian Wilson loop $W^{\mathrm{Abel}}(C)$ is defined as the Wilson loop for the Abelian link variables as
\begin{align}
  W^{\mathrm{Abel}}(C) := \tr\left\{\prod_{\braket{x,\mu}\in C} u_{x,\mu}\right\},
\end{align}
which is considered as the ``dominant'' part of the Wilson loop.
The monopole contribution is extracted from the Abelian Wilson loop by applying the Toussaint-DeGrant procedure \cite{DT80}.

\section{Higher representations}
If the Abelian projection is applied naively to the static sources in a higher representation, the correct infrared behavior of the static potential is not reproduced.
For example in the adjoint representation of $SU(2)$, the limit of the Abelian Wilson loop average as the loop size approaches infinity seems to be $1/3$ in the numerical simulation \cite{Poulis96}.
In this case, we cannot extract the static potential because that is extracted from the exponential decay for $T$ of the average of the rectangular Wilson loop with length $T$ and width $L$.
In the spin-$3/2$ representation, the string tension extracted from the Abelian Wilson loop is the same as one for the fundamental representation \cite{DFGO96}, which is different from the correct one.

In this talk we claim that we should use a different operator instead of the naive Abelian Wilson loop $W^{\mathrm{Abel}}_R$ in a representation $R$.
The operator is defined as follows.
We parameterize the untraced Abelian Wilson loop $w_C$ with $\phi_a\in\mathbb R$ ($a=1,\ldots r$) as
\begin{align}
  w_C = \exp\left( i\sum_{a=0}^r \phi_a H_a \right),
\end{align}
where $r$ is the rank of the group and $H_a$ is a Cartan generator.
In this parametrization the naive Abelian Wilson loop $W^{\mathrm{Abel}}_R$ in a representation $R$ is written as
\begin{align}
  W^{\mathrm{Abel}}_R = \frac1{D_R}\sum_{\bm\mu\in \Delta_R}d_\mu\exp\left( i\sum_{a=0}^r \mu_a\phi_a \right),
\end{align}
where $D_R$ is the dimension of $R$, $\Delta_R$ is the set of the weights of $R$, $\mu_a$ is the $a$-th component of a weight $\bm\mu$ and $d_\mu$ is the dimension of the weight space with weight $\bm\mu$, which satisfies $D_R = \sum_{\bm\mu\in \Delta_R} d_\mu$.
Instead of this, we propose the operator defined by
\begin{align}
  \tilde W^{\mathrm{Abel}}_R := \frac1{D^h_R}\sum_{\bm\Lambda\in \Delta^h_R}\exp\left( i\sum_{a=0}^r \Lambda_a\phi_a \right), \label{tildeW}
\end{align}
where $\Delta^h_R$ is the set of the weights equivalent to the highest weight of $R$ under the action of the Weyl group and $D^h_R$ is the number of elements in $\Delta^h_R$.
This operator \cref{tildeW} can be written by using the untraced Abelian Wilson loop $w_C := \prod_{\braket{x,\mu}\in C}u_{x,\mu}$.
For $SU(2)$, in the spin-$J$ representation,
\begin{align}
  \tilde W^{\mathrm{Abel}}_J(C) = \frac12\tr ( w_C^{2J} ) . \label{su2}
\end{align}
For $SU(3)$, in the representation with the Dynkin index $[m_1,m_2]$,
\begin{align}
 \tilde W^{\mathrm{Abel}}_{[m_1,m_2]}(C) =
\begin{cases}
   \frac16\left(\tr (w_C^{m_1}) \tr(w_C^{*m_2}) - \tr(w_C^{m_1-m_2}) \right)  &\text{for}\quad m_1\geq m_2\neq 0
\\
   \frac16\left(\tr (w_C^{m_1}) \tr(w_C^{*m_2}) - \tr(w_C^{*m_2-m_1}) \right)  &\text{for}\quad m_2\geq m_1\neq 0
\\
   \frac13\tr (w_C^{m_1})  &\text{for}\quad m_2=0 
\\
   \frac13\tr (w_C^{*m_2})  &\text{for}\quad m_1=0 ,
  \end{cases}
\label{su3}
\end{align}
where $w_C^0 = \bm 1$.

\subsection*{The origin of the operator \cref{tildeW}}
Here we give the reason why the proposed operator \cref{tildeW} could reproduce the correct behavior.
First we consider the question: why cannot the Abelian Wilson loop in a higher representation reproduce the correct behavior of the Wilson loop?
To answer this, we introduce the distribution function of the untraced Wilson loop for a rectangular loop with length $T$ and width $L$ as
\begin{align}
  P(W;L,T) := \int DU \delta_G(W, \prod_{l\in C_{L,T}}U_l) e^{-S[U]}, \label{dis}
\end{align}
where $DU$ is the functional integral measure for the link variables $U_{x,\mu}$, $\delta_G(\bullet,\bullet)$ is the delta function on the group $G$, $C_{L,T}$ is a rectangular loop with length $T$ and width $L$, and $S[U]$ is the action.
This has the information of the average of the Wilson loop in an arbitrary representation. 
Conversely $P(W;L,T)$ can be written by using the averages of the Wilson loops in the whole representations through the character expansion. 
On the other hand, the distribution function for the untraced Abelian Wilson loop can be written as
\begin{align}
   P_{\mathrm{Abel}}(w;L,T) = \int DU \delta_H(w,\prod_{l\in C_{L,T}}u_l) e^{-S[U]}, \label{abeldis}
\end{align}
where $u_l$ is the Abelian link variable defined in \cref{sec:Abel} and $\delta_H(\bullet,\bullet)$ is the delta function on the Cartan subgroup $H$.
In the following we see that the asymptotic behaviors of the Wilson loop average and the Abelian Wilson loop average can be different even if the asymptotic behaviors of \cref{dis,abeldis} are the same.
The average of the Wilson loop in every representation except the trivial one approaches zero in the large $T$ limit because $\braket{W_R(L,T)} \simeq C(L)e^{-V_R(L)T} \rightarrow 0$ ($T\rightarrow \infty)$.
Hence the distribution $P(W;L,T)$ of the untraced Wilson loop approaches the uniform distribution $P(W;L,T) \rightarrow 1$ as $T$ tends to infinity.
On the other hand, if we assume that the distribution $P_{\mathrm{Abel}}(w;L,T)$ of the untraced Abelian Wilson loop also approaches the uniform distribution, the average of the Abelian Wilson loop in a representation does not necessarily approaches zero as $T$ tends to infinity.
As an example, we consider $SU(2)$ case.
To see the difference between the Wilson loop and the Abelian Wilson loop, we explicitly calculate the asymptotic behavior of the Wilson loop average in the representation $R$ by using the fact that $P(W;L,T)$ approaches the uniform distribution.
We parameterize the untraced Wilson loop as
\begin{align}
  W &= \begin{pmatrix}
    \cos\theta + i\sin\theta\cos\theta_1 & \sin\theta\sin\theta_1e^{i\theta_2}\\
    -\sin\theta\sin\theta_1e^{-i\theta_2} & \cos\theta -i\sin\theta\cos\theta_1
  \end{pmatrix} \in SU(2). 
\end{align}
The only gauge invariant parameter is $\theta$ and therefore we can simply write $P(W;L,T) = P(\theta;L,T)$.
In this parameterization the Haar measure on $SU(2)$ is
\begin{align}
  \int dW = \frac1{2\pi^2}\int_0^\pi d\theta \int_0^\pi d\theta_1 \int_0^{2\pi} d\theta_2 \sin^2\theta \sin\theta_1.
\end{align}
Thus we calculate the asymptotic behavior of the average of the Wilson loop in the spin-$J$ representation as
\begin{align}
  \braket{W_J(L,T)} &= \int dW P(W;L,T)\frac1{D_J}\tr_J W = \frac2\pi\int_0^\pi d\theta \sin^2\theta P(\theta;L,T)\frac1{2J+1}\sum_{k=0}^{2J}e^{2i(k-J)\theta} \notag\\
  &\overset{T\rightarrow\infty}{\longrightarrow} \frac2\pi\int_0^\pi d\theta \sin^2\theta \frac1{2J+1}\sum_{k=0}^{2J}e^{2i(k-J)\theta} =0. \label{Wj}
\end{align}
Next we consider the Abelian Wilson loop.
We parameterize the untraced Abelian Wilson loop as
\begin{align}
  w = \operatorname{diag}(e^{i\theta},e^{-i\theta}) \in U(1).
\end{align}
In this parametrization we can write $P_{\mathrm{Abel}}(w;T,L) = P_{\mathrm{Abel}}(\theta;T,L)$.
If $P_{\mathrm{Abel}}(\theta;T,L) \rightarrow 1$ ($T\rightarrow \infty$), then the asymptotic behavior of the Abelian Wilson loop average in the spin-$J$ representation is calculated by using the Haar measure on $U(1)$ as
\begin{align}
  \braket{W^{\mathrm{Abel}}_J(L,T)} &\overset{T\rightarrow\infty}{\longrightarrow} \frac1{2\pi}\int_0^{2\pi} d\theta \frac1{2J+1}\sum_{k=0}^{2J} e^{2i(k-J)\theta} 
  = \begin{cases}
    0 & \text{if $J$ is a half-integer}\\
    \frac1{2J+1} & \text{if $J$ is an integer},
  \end{cases} \label{WAj}
\end{align}
which is clearly different from \cref{Wj}.
This result is consistent with the numerical results for $J=1/2,1,3/2$ \cite{DFGO96,Poulis96}.
The difference between \cref{Wj,WAj} is due to the factor $\sin^2\theta$ in \cref{Wj}, which comes from the Haar measure on $SU(2)$.
We can say that this is the reason why the Abelian Wilson loop average in a higher representation does not reproduce the correct behavior.
This consideration leads to the hypothesis that the difference between the behaviors of the Wilson loop average and the Abelian Wilson loop average is \textit{only} due to the difference between the measure on $SU(2)$ and $U(1)$.
We can correct this difference of the measure by modifying the operator calculated.
The difference between the measure on $SU(2)$ and $U(1)$ is the factor $\sin^2\theta$ and therefore in the hypothesis the following approximation holds,
\begin{align}
 \braket{\sin^2\hat\theta(L,T)W^{\mathrm{Abel}}_J(L,T)} \simeq C_J(L)e^{-V_J^0 T}\braket{W_J(L,T)}   \label{approx}
\end{align}  
for large $L$ and $T>>L$, where $e^{i\hat\theta(L,T)}$ is an eigenvalue of the untraced Abelian Wilson loop, $V_J^0$ is a real constant which corresponds to a shift of the static potential and $C_J(L)$ does not depend on $T$ and therefore does not contribute to the static potential.
Actually, this hypothesis is equivalent to the statement that the proposed operator \cref{tildeW} behaves correctly if we assume that the Wilson loop average falls off faster in the size of the loop as its representation becomes higher.
The proof is sketched as follows, see \cite{MSKK} for details.
The right hand side \cref{approx} can be expressed by using the proposed operator \cref{tildeW} as
\begin{align}
  \braket{\sin^2\hat\theta\, W^{\mathrm{Abel}}_J} 
  &= \frac1{2(2J+1)}(\braket{\tilde W_J^{\mathrm{Abel}}} - \braket{\tilde W_{J+1}^{\mathrm{Abel}}}). \label{rhs}
\end{align}
Due to the Riemann-Lebesgue lemma, 
we can solve \cref{rhs} for $\braket{\tilde W_J^{\mathrm{Abel}}}$ as
\begin{align}
  \braket{\tilde W_J^{\mathrm{Abel}}} = 2\sum_{k=2J}^{\infty}(k+1)\braket{\sin^2\hat\theta\,W_{k/2}^{\mathrm{Abel}}}.
\end{align}
If the hypothesis \cref{approx} holds, then we obtain
\begin{align}
  \braket{\tilde W_J^{\mathrm{Abel}}(L,T)} \simeq 2\sum_{k=2J}^{\infty}(k+1)C_{k/2}(L)e^{-V_{k/2}^0T}\braket{W_{k/2}(L,T)}.
\end{align}
If we assume that the Wilson loop average falls off faster as its representation becomes higher, we obtain
\begin{align}
  \braket{\tilde W_J^{\mathrm{Abel}}(L,T)} \simeq 2(2J+1)C_J(L)e^{-V_J^0T}\braket{W_J(L,T)}, \label{approx2}
\end{align}
which is nothing but our proposal.
Conversely, by assuming \cref{approx2}, substituting it into \cref{rhs} and assuming that the Wilson loop average falls off faster as its representation becomes higher, we obtain \cref{approx}.
Thus \cref{approx,approx2} are equivalent.

\begin{figure}[t]
\centering
\includegraphics[width=0.35\hsize]{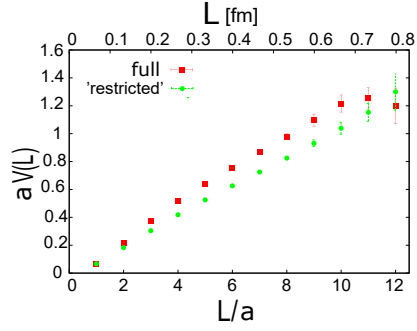}
\caption{
The static potential between the sources in the adjoint representation of $SU(2)$ using \cref{su2} for $J=1$ and for comparison the full Wilson loop average in the adjoint representation. The result is consistent with that of \cite{Poulis96,CHS04} where the same quantity is calculated.
}
\label{fig:su2}
\end{figure}

In view of \cref{approx}, our proposal means that we can correct the behavior of the Abelian Wilson loop by modifying it to include the effect of the difference between the Haar measure on the group and the Cartan subgroup.
The similar argument is possible in case of an arbitrary compact group \cite{MSKK}.

\section{Numerical result} \label{num_result}

\begin{figure}[t]
\begin{minipage}{0.33\hsize}
\centering
\scalebox{0.77}{\includegraphics{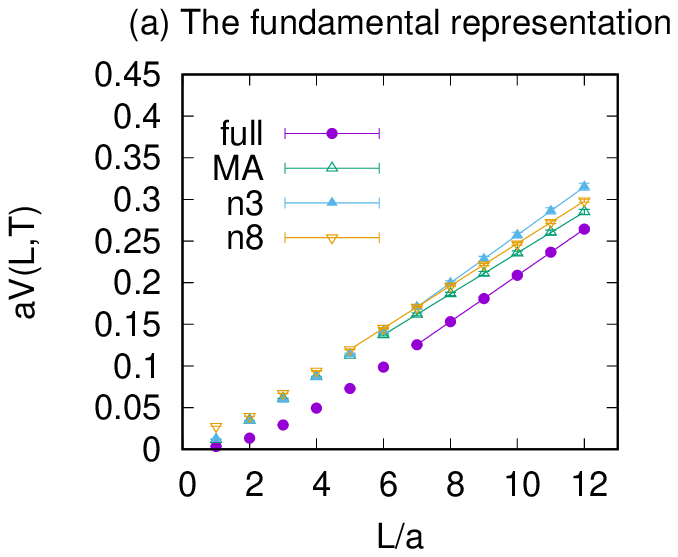}}
\end{minipage}
\begin{minipage}{0.33\hsize}
\centering
\scalebox{0.77}{\includegraphics{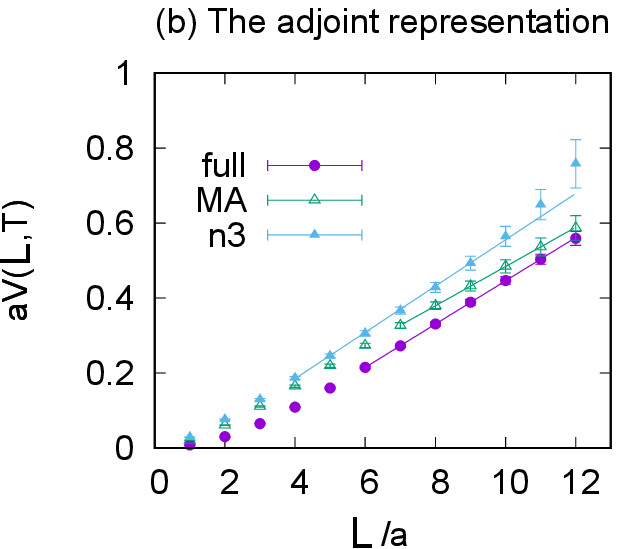}}
\end{minipage}
\begin{minipage}{0.33\hsize}
\centering
\scalebox{0.77}{\includegraphics{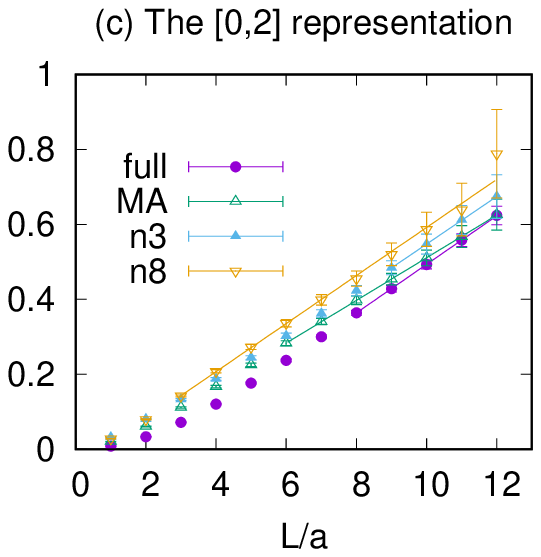}}
\end{minipage}
\caption{
The static potential $\braket{V(R,T=8)}$ between the sources in (a) the fundamental $[1,0]$, (b) the adjoint $[1,1]$, and  (c) the sextet $[0,2]$ representations of $SU(3)$ calculated using \cref{su3} in comparison with the full Wilson loop average.
The legends, MA, n3 and n8 represents the measurements by using the corresponding reduction conditions \cref{MA,n3,n8} respectively.
}
\label{fig:su3}
\end{figure}

\begin{table}[t]
\centering
\caption{
The string tensions in the lattice unit in the $SU(3)$ case:  the string tensions obtained under the MA gauge (\cref{MA}), and other gauges n3 (\cref{n3}) and n8 (\cref{n8}), in comparison with the full string tension.
}
\begin{tabular}{c||c|c|c|c} \hline
&full & MA & n3 & n8 \\\hline\hline
fundamental & $0.02776(2)$ & $0.02458(1)$ & $0.02884(3)$ & $0.02544(3)$ \\\hline
adjoint & $0.0576(1)$ & $0.0522(1)$ & $0.062(1)$ &- \\\hline
[0,2] & $0.0647(1)$ & $0.05691(9)$ & $0.0635(2)$ & $0.0641(6)$ \\\hline 
\end{tabular}
\label{tab}
\end{table}

In order to support our claim that the dominant part of the Wilson loops in higher representation is given by \cref{tildeW},
we check numerically whether the string tension extracted from \cref{su2,su3} reproduce the full string tension or not.

For this purpose we set up the gauge configuration for the standard Wilson action 
at $\beta = 2.5$ on the $24^4$ lattice for SU(2) and at $\beta=  6.2$  on the $24^4$ lattice for SU(3).
For SU(2) case, we prepare 500 configurations   every 100 sweeps after 3000 thermalization by using the heatbath method.
For SU(3) case, we prepare 1500 configurations  evey  50  sweeps after 1000 thermalization   by using pseudo heatbath method with over-relaxation algorithm (20 steps per sweep).
In the measurement of the Wilson loop average we apply the APE smearing technique for $SU(3)$ case and the hyper-blocking for $SU(2)$ case to reduce noises and the exciting modes.
The number of the smearing is determined so that the ground state overlap is enhanced \cite{BSS95}.
We have calculated the Wilson loop average $W(L,T)$ for a rectangular loop with length $T$ and width $L$ to derive the potential $V(L,T)$ through the formula
$V(L,T) := -\log(W(L,T+1)/W(L,T)$.

The result for $SU(2)$ is given in \cref{fig:su2} for the adjoint representation.
The results for $SU(3)$ are given in \cref{fig:su3}  (a) for the fundamental representation $[1,0]$, in \cref{fig:su3} (b) for the adjoint representation $[1,1]$ and in \cref{fig:su3} (c) for the  sextet representation $[0,2]$.
\Cref{tab} shows the string tensions which are extracted by fitting the data with the linear potential. 
The string tensions extracted from the "dominant" operators we proposed reproduce nearly equal to or more than $80\%$ of the full string tension.
These results indicate that the proposed operators give actually the dominant part of the Wilson loop average.

\section{Conclusion} \label{conc}

In this talk, we have proposed a solution for the problem that the correct behavior of the Wilson loop in higher representations cannot be reproduced if the restricted part of the Wilson loop is naively extracted by applying the Abelian projection or the field decomposition in the same way as in the fundamental representation.
We have proposed the prescription to construct the operator suitable for this purpose. 
We have performed numerical simulations to show that this prescription works well in the adjoint rep.\ $\bf{3}$ for $SU(2)$ color group, and the adjoint rep.\ $[1,1]=\bf{8}$ and the sextet rep.\ $[0,2]=\bf{6}$ for $SU(3)$ color group.
Further studies are needed in order to establish the magnetic monopole dominance in the Wilson loop average for higher representations, supplementary to the fundamental representation for which the magnetic monopole dominance was established.

\section*{Acknowledgement}

This work was supported by Grant-in-Aid for Scientific Research, JSPS KAKENHI Grant Number (C) No.15K05042.
R. M. was supported by Grant-in-Aid for JSPS Research Fellow Grant Number 17J04780. 
The numerical calculations were in part supported by the Large Scale Simulation Program No.16/17-20(2016-2017) of High Energy Accelerator Research Organization (KEK), and were performed in part using COMA(PACS-IX)  at the CCS, University of Tsukuba.

\end{document}